\documentstyle[emulateapj]{article}

\slugcomment{Accepted for publication in ApJ Letters}
\lefthead{Burke et al.}
\righthead{BCG Evolution and Cluster Selection}

\newcommand{\ho}[1]{{\rm H$_0 = #1$~km\,\persec\,\permpc}}

\newcommand{\lx}{\mbox{$L_{x}$}}

\newcommand{\lmineds}{2.3}

\newcommand{\nclus}{76}   
\newcommand{\abscal}{0.05}

\newcommand{\zf}{\mbox{$z_f$}}

\newcommand{\persec}{\mbox{s$^{-1}$}}
\newcommand{\permpc}{\mbox{Mpc$^{-1}$}}
\newcommand{\lumin}{\mbox{erg \persec}}

\newcommand{\centre}{center}
\newcommand{\colour}{color}
\newcommand{\favoured}{favored}

\newcommand{\programmes}{programs}

\newcommand{\semianalytic}{semi-analytic}
\newcommand{\semianalytical}{semi-analytical}
\newcommand{\xray}{X-ray}

\newcommand{\eg}{\mbox{e.g.}}
\newcommand{\etal}{\mbox{et al.}}

\newcommand{\cmfirst}{Collins \& Mann~\markcite{cm98}(1998; hereafter CM98)}
\newcommand{\pcmmain}{\protect\markcite{cm98}CM98}
\newcommand{\cmmain}{\markcite{cm98}CM98}

\newcommand{\abkfirst}{Arag\'{o}n-Salamanca, Baugh \& Kauffmann \markcite{abk98}(1998; hereafter ABK98)}
\newcommand{\pabkmain}{\protect\markcite{abk98}ABK98}
\newcommand{\abkmain}{\markcite{abk98}ABK98}

\newcommand{\deprop}{De Propris \etal~\markcite{depropris99}(1999)}

\newcommand{\lynam}{Lynam \etal~\markcite{pdl99}(1999)}

\begin{document}

\title{Cluster Selection and the Evolution of Brightest Cluster Galaxies}

\author{
D. J. Burke\altaffilmark{1,2,5},
C. A. Collins\altaffilmark{1,5},
R. G. Mann\altaffilmark{3,4,5}}

\altaffiltext{1}{Astrophysics Research Institute, Liverpool John Moores University,
12 Quays House, Egerton Wharf, Birkenhead, CH41 1LD, UK}
\altaffiltext{2}{Institute for Astronomy, University of Hawaii,
2680 Woodlawn Drive, Honolulu, HI 96822}
\altaffiltext{3}{Astrophysics Group, Blackett Laboratory, 
Imperial College, Prince Consort Road, London, SW7 2AZ, UK}
\altaffiltext{4}{Institute for Astronomy, University of Edinburgh, 
Royal Observatory, Blackford Hill, Edinburgh, EH9 3NJ, UK}
\altaffiltext{5}{Visiting Astronomer at the 
NASA Infrared Telescope
Facility, which is operated by the University of
Hawaii under contract from the National Aeronautics
and Space Administration.}

\begin{abstract}
The K-band Hubble diagram of Brightest Cluster Galaxies (BCGs) is presented
for a large, \xray\ selected cluster sample extending out to $z = 0.8$.
The controversy over the degree of BCG evolution is shown to be due to
sample selection,
since the BCG luminosity depends upon the cluster environment.
Selecting only the most \xray\ luminous clusters produces a BCG sample 
which 
shows, under the assumption of an Einstein-de Sitter cosmology, 
significantly less mass growth than that predicted by current 
\semianalytic\ galaxy formation models, and significant evidence of 
any growth only if the
dominant stellar population of the BCGs formed relatively recently ($z \leq 2.6$).

\end{abstract}

\keywords{Galaxies: clusters: general --- galaxies: elliptical and lenticular, 
cD --- galaxies: evolution --- galaxies: formation}

\section{INTRODUCTION}
\label{intro}

The majority of stars in giant ellipticals 
found in the cores of rich galaxy clusters are old;
photometric and spectroscopic studies of cluster galaxies out to
$z \approx 1$ suggest a formation redshift, \zf, greater than 2,
with little variation within a cluster, 
and that secondary bursts of star formation account for
a small fraction of the stellar mass
(\eg\ 
Arag\'{o}n-Salamanca \etal~\markcite{aecc93}1993;
Ellis \etal~\markcite{ellis97}1997;
Stanford, Eisenhardt \& Dickinson~\markcite{sed98}1998;
van Dokkum \etal~\markcite{vd98}1998;
Poggianti \etal~\markcite{morphs99}1999).

However, to understand the process of galaxy formation it is necessary to
know where the stars were formed as well as when.
In the traditional view of early-type galaxy formation---a
``monolithic'' collapse at high redshift 
(\eg\ Eggen, Lynden-Bell \& Sandage~\markcite{els62}1962;
Larson~\markcite{larson69}1969)---all the stars
were formed in situ, in direct contrast to the merger-driven
growth of galaxies predicted by 
\semianalytical\ models for hierarchical cosmologies, such as CDM
(\eg\ Kauffmann \& White~\markcite{kw93}1993;
Baugh, Cole \& Frenk~\markcite{bcf96}1996).
Since the ages of the stars are similar in both scenarios,
it is the change in mass with look-back time that
separates the two 
pictures
observationally.
The current data are inconclusive;
for example,
the hierarchical models are \favoured\ by the enhanced
merger fraction seen in the $z=0.8$ cluster MS1054.4-0321
(van Dokkum \etal~\markcite{v99}1999), whilst
\deprop\ show 
no evidence for mass evolution of bright ellipticals in clusters
out to $z \approx 1$.
In general it is difficult to follow the evolutionary history of 
ellipticals since selection methods can seriously bias the samples,
c.f. the discussions of progenitor bias in van Dokkum \& Franx~\markcite{vf96}(1996),
the effect of preferential selection of the most massive objects at
each epoch in Kauffmann \& Charlot~\markcite{kc98}(1998),
and the use of \colour\ selection in Jimenez \etal~\markcite{jimenez99}(1999).

One approach to minimising such problems is to study the evolution of 
a particular class of ellipticals---brightest cluster galaxies (BCGs)---because
of their unique location, close to the \centre\ of the cluster's
potential well.
BCGs do not appear to be drawn from the same luminosity function
as other cluster galaxies (\eg\ Dressler~\markcite{dressler78}1978), 
which suggests that they have a distinct formation history.
Knowledge of BCG evolution can therefore provide different
constraints on galaxy formation models to studies
of the general cluster population.

The K-band Hubble diagram for BCGs has recently been extended
to $z \approx 1$ by 
both \cmfirst\ and 
\abkfirst: these
observations provide the opportunity to measure the luminosity
evolution of BCGs since evolutionary and pass-band 
corrections are insensitive to the recent star-formation history of
a galaxy at near-IR wavelengths (\eg\ Bershady~\markcite{bershady95}1995;
Madau, Pozzetti, \& Dickinson~\markcite{madau98}1998).
The conclusions drawn are contradictory,
despite the use of the same cosmology and a common assumption that
the stellar populations of BCGs are old and passively evolving;
\cmmain\ assert that the stellar populations of BCGs in the most massive clusters
have not grown significantly since $z \approx 1$,
whilst \abkmain\ argue that their results are in good agreement with 
the mass increase of BCGs---by a factor of four in an
Einstein-de Sitter cosmology---predicted by 
\semianalytical\ 
models over the same redshift range.
The two samples have almost no overlap---\cmmain\ having used
an \xray\ selected cluster catalogue whilst
\abkmain\ used a heterogeneous compilation that was mainly optically selected---and
it is the aim of the present work to show that the results
can be reconciled by considering the properties of the 
clusters in the two samples.
Section~\ref{data} describes the BCG sample used---
an extension of that of \cmmain---and
the reduction methods employed, whilst
section~\ref{results} presents the results of the analysis
and a comparison to those of \abkmain.
Throughout this letter
an Einstein-de Sitter cosmology with \ho{50} is assumed,
and \xray\ luminosities (\lx) are quoted for the 0.3--3.5~keV pass band.

\section{DATA}
\label{data}

\subsection{Sample}
\label{data:sample}

The data presented here comprise K-band observations of \nclus\ BCGs.
This sample, which incorporates that of \cmmain,
spans a redshift range of 0.05 to 0.83, and is drawn from the following
\xray\ selected cluster catalogues:
the Einstein EMSS (Gioia \& Luppino \markcite{gl94}1994; 
Nichol \etal\ \markcite{n97}1997; Henry \markcite{h99}1999),
the Southern and Bright SHARC catalogues 
(Burke \etal\ \markcite{b97}1997; 
Romer \etal\ \markcite{bsharc}2000),
and the ROSAT NEP Survey (Henry \etal\ \markcite{nep97}1997).
\xray\ selection is to be preferred, since both \xray\ luminosity and 
\xray\ temperature should be more closely related to cluster
mass than optical richness.

The \xray\ luminosity-redshift coverage of the cluster sample is shown
by the circles in Figure~1. 
The additional symbols show those clusters from \abkmain\ with a 
measured \xray\ flux or upper limit,
except for Cl~2155+0334 (also known as Cl~2157+0347), which
has been removed because photometric and
spectroscopic observations show no evidence for a cluster
(Thimm \& Belloni~\markcite{tb94}1994; Oke, Postman \& 
Lubin~\markcite{opl98}1998), and Cl~0016+16, since it is in both samples.
The difference in \xray\ luminosity coverage at $z > 0.5$
for the two samples is striking;
the implications of this are discussed in section~\ref{results}.

\vspace{2mm}
\begin{center}
%
\plotfiddle{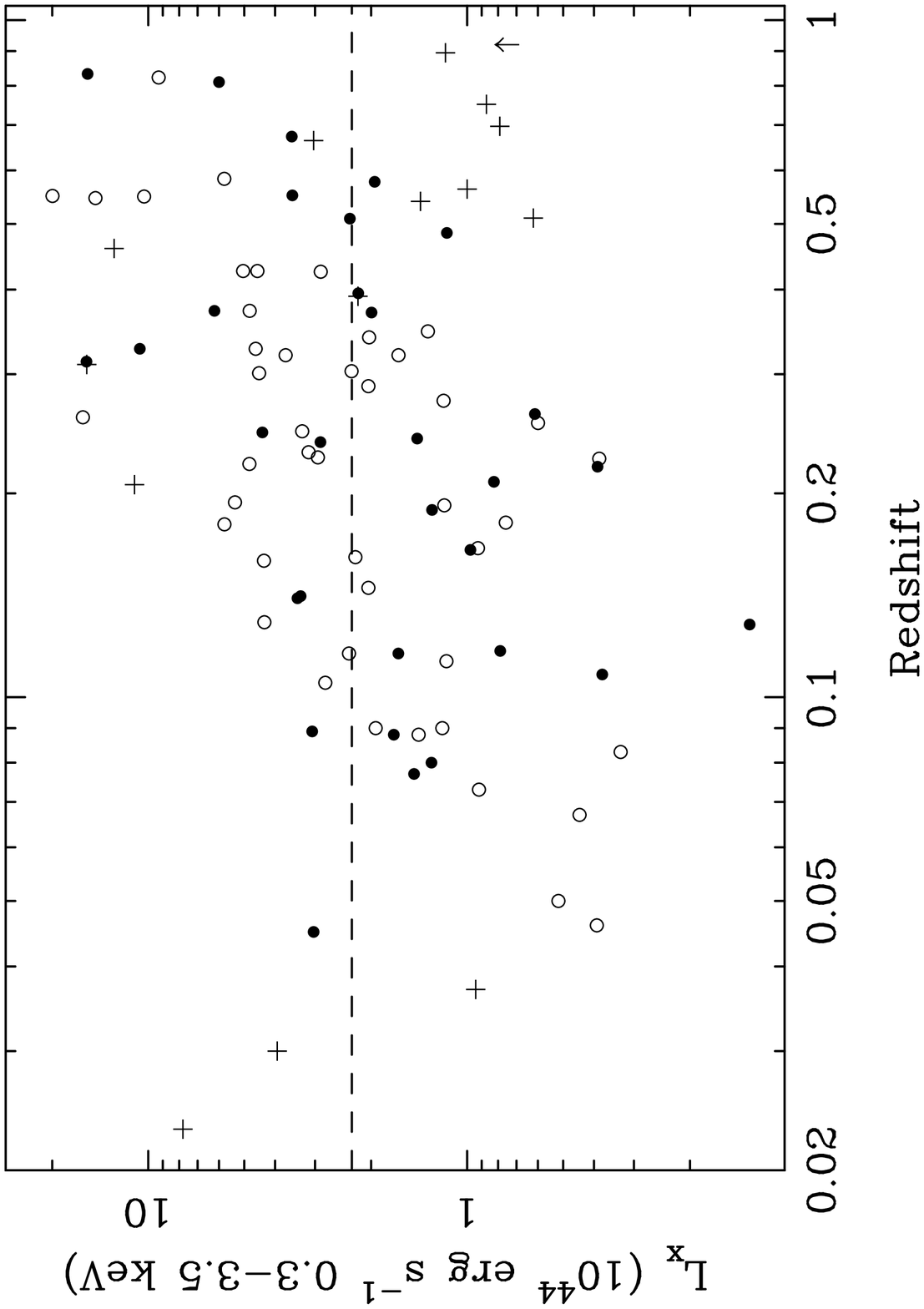}{2in}{-90}{32}{32}{-120}{190}
\begin{minipage}{8.75cm}
\small\parindent=3.5mm
{\sc Fig.}~1.---Cluster \xray\ luminosity as a function of redshift.
The circles indicate the BCG sample presented here; 
open for clusters used 
in \pcmmain\ and filled for the new observations.
Clusters from \pabkmain\ with measured \xray\ fluxes are shown
as plus ($+$) symbols and the arrow ($\uparrow$) symbol 
represents the $3\sigma$ upper limit for Cl~1603+4329.
The dashed line, at $\lx = \lmineds \times 10^{44}$~\lumin, 
shows the luminosity used by
\pcmmain\ to separate their sample into high- and low-\lx\ clusters.
\par
\end{minipage}
\end{center}
\vspace{3mm}

The K-band BCG observations were made using the IRCAM3 and UFTI cameras on the
UKIRT and NSFCAM on the IRTF, with some of the
data being provided by the service \programmes\ of both telescopes.
IRCAM3 and NSFCAM are 256x256 pixel InSb devices with a field of
view close to 70\arcsec\ by 70\arcsec\ (the IRCAM3 and NSFCAM
pixel scales are 0.281 and 0.3~\arcsec\,pixel$^{-1}$
respectively), and UFTI is a 1024x1024 pixel HgCdTe array with
a pixel scale of 0.091~\arcsec\,pixel$^{-1}$,
giving a field of view of 92\arcsec\ by 92\arcsec.
The observing strategy is the same as presented in \cmmain:
the BCGs were imaged using a jitter pattern and
separate sky exposures were taken 
for those objects which filled the field of view.

\subsection{Reduction}
\label{data:reduction}

The data reduction system improves upon that presented in 
\cmmain, and incorporates elements from the methods described
in Stanford, Eisenhardt \& Dickinson~\markcite{sed95}(1995) 
and Hall, Green \& Owen~\markcite{hall98}(1998).
An outline is presented below as the method 
will be fully described in a later paper.

The individual frames were masked for bad pixels, dark subtracted
and divided by the exposure time.
A flat field image was created by median combination of the 
object images---or separate sky exposures if these were available---and
applied to the object frames. 
Masking of cosmic ray events was performed 
on the flattened images before they were mosaiced together,
which completed the processing of those objects with sky exposures.
Otherwise the mosaic---which is substantially deeper than 
the individual exposures---was used to create an object mask,
which was then applied to the individual images before they were
median-combined to form a flat.
The flattened exposures were then processed as above to create the
final image.

Observations of stars from the UKIRT faint standards 
list (Casali \& Hawarden~\markcite{ukirt-fs}1992)
were used to calibrate the photometry onto the UKIRT system
assuming an extinction of 0.088 mag airmass$^{-1}$, 
the median value for K-band observations at Mauna Kea.
Comparisons of the results from repeat observations, both
within and between observing runs, show that the magnitudes
agree to \abscal\ mag.

Aperture magnitudes were measured using a 50~kpc diameter aperture
and have been corrected for Galactic absorption using the
maps of Schlegel, Finkbeiner, \& Davis~\markcite{schlegel98}(1998):
the correction is small, mostly being less than 0.05~mag, but
reaching 0.1~mag in several cases.
The position of the aperture was chosen so as to maximise the 
flux contained within it whilst remaining close to the \centre\ of
the cluster \xray\ emission.
Those pixels contaminated by stars and obvious non-cluster
galaxies were excluded from the calculation, being replaced by values 
chosen from regions at the same distance from the aperture \centre.
No attempt has been made to remove flux due to other cluster galaxies
falling within the aperture, and so the results are
directly comparable to those of \abkmain.

\section{RESULTS}
\label{results}

The K-band Hubble diagram for the two BCG samples is shown in 
Figure~2. The lines show model predictions
calculated using the GISSEL96 code (Bruzual \& Charlot \markcite{bc}1993),
for a solar-metallicity stellar population with a Salpeter initial mass function:
the solid line indicates a no-evolution model for a 10~Gyr old
stellar population, whereas the other lines are for stellar populations 
which form in an instantaneous burst of star formation at a single 
epoch---$\zf = 2$ for the dashed line and $\zf = 5$ for the dotted line---and then 
evolve passively.
The models have been normalised to match the low-redshift, X-ray selected, 
BCG sample of \lynam, following the method used in \abkmain, 
assuming a growth curve, $d\,\log{L}$/$d\,\log{r}$, of 0.7
for the aperture corrections and a \colour\ of $R-K = 2.6$.


\vspace{2mm}
\begin{center}
%
\plotfiddle{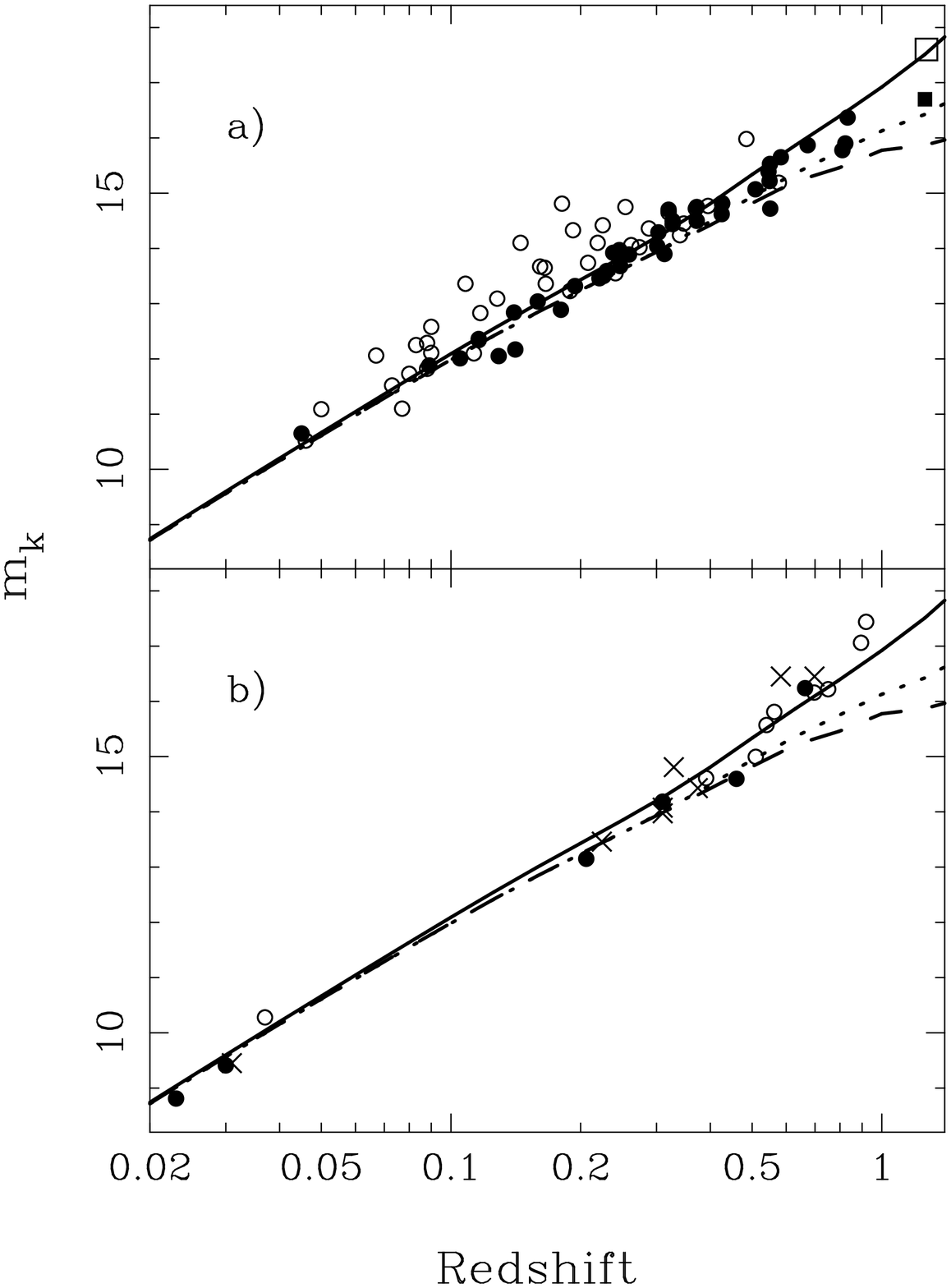}{4.2in}{0}{42}{42}{-120}{-10}
\begin{minipage}{8.75cm}
\small\parindent=3.5mm
{\sc Fig.}~2.---Magnitude-redshift relation for brightest cluster galaxies
in the observed K band.
Filled and open symbols represent those 
BCGs in high- and low-\lx\ clusters respectively; the
division is as in Figure~1.
The circles in the top panel indicate the sample presented here,
whilst the squares are for the BCGs in the
two $z=1.3$ clusters from Rosati \etal~\protect\markcite{rosati99}(1999),
where the magnitudes have been measured within 50~kpc diameter apertures
(P. Rosati~1999, private communication).
The bottom panel shows the sample of \pabkmain, where 
the crosses are for those clusters without a measured \xray\ flux.
The no-evolution prediction, assuming a 10~Gyr old
stellar population, is shown by the solid line;
passive-evolution models, in which the stars form at a single
epoch, are shown as dashed ($\zf = 2$)
and dotted ($\zf = 5$) lines.
\par
\end{minipage}
\end{center}
\vspace{3mm}

The result remains qualitatively the same as Figure~6 of \cmmain; BCGs
in high-\lx\ clusters form a homogeneous population which is
brighter, and has a smaller scatter, than that of low-\lx\ clusters.
This can be more clearly seen in Figure~3,
which shows the scatter around the model predictions
as a function of cluster \xray\ luminosity.
It is this relationship between BCG and cluster properties that leads
to the contradictory conclusions of \cmmain\ and \abkmain:
out of the eleven $z > 0.5$ clusters in the latter sample, nine have
\xray\ flux measurements or upper limits, with all but two of these having
a low \xray\ luminosity (Figure~1).
It is unsurprising that these clusters are not similar
to rich, local clusters, as they
were discovered on the basis of their optical properties
(\eg\ Castander \etal~\markcite{castander94}1994;
Holden \etal~\markcite{holden97}1997).
The squares in Figure~2 represent the 
BCGs in the two $z = 1.3$ clusters discussed by
Rosati \etal~\markcite{rosati99}(1999): the high-\lx\ cluster (solid square)
was discovered by means of its \xray\ emission,
whereas the low-\lx\ cluster (open square) was detected by its galaxy population. 
Although based on only two points, this suggests that the correlation 
with environment holds at this redshift. 


The \semianalytic\ models discussed in \abkmain\ predict a factor of 
$\sim $4--5 increase in the stellar masses of BCGs in
massive clusters since $z=1$, for an Einstein-de Sitter universe.
To test whether the data presented here supports this level of evolution, 
a correlation between redshift and the BCG residuals
($\Delta m_k$, \eg\ Figure~4)
has been sought.
Passive-evolution models with $\zf=2$ and $\zf=5$ have been used to
calculate the residuals---since they provide a conservative range for the 
formation epoch of massive cluster ellipticals 
(\eg\ Ellis \etal~\markcite{ellis97}1997)---and
separate fits made to the high- and low-\lx\ cluster subsamples.
Since the form of any evolution is unknown a priori, a non-parametric
rank-order statistic---Kendall's $\tau$---was used; it also has the advantage that
it is insensitive to the choice of normalisation adopted for the Bruzual
\& Charlot models.
All save one of the fits showed no significant ($>3\sigma$) evidence for
evolution; the exception, at a significance of $3.6\sigma$, was the
high-\lx\ subsample with $\zf=2$.
To find the maximum formation epoch that is still compatible with 
evolution of the high-\lx\ subsample, \zf\ was increased 
from 2 until the correlation significance dropped below $3\sigma$.
Evolution is found only if the stars formed recently ($\zf \leq 2.6$).

\vspace{2mm}
\begin{center}
%
\plotfiddle{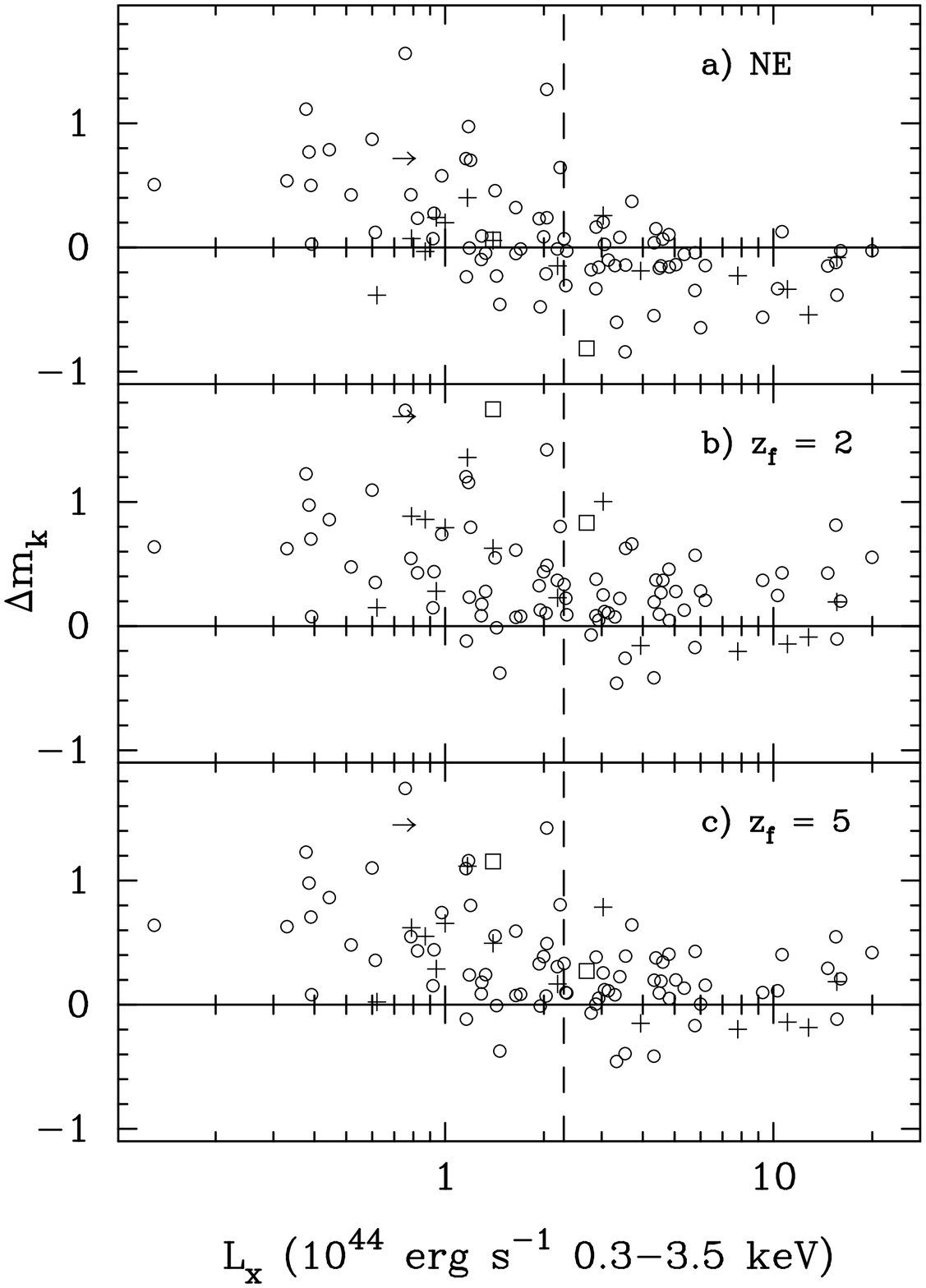}{4.1in}{0}{42}{42}{-120}{-10}
\begin{minipage}{8.75cm}
\small\parindent=3.5mm
{\sc Fig.}~3.---Residuals about the model predictions, defined as 
$\Delta m_k = m_{\rm BCG} - m_{\rm model}$, as a function of 
cluster \xray\ luminosity.
The BCGs presented here are shown as circles, the
sample of \pabkmain\ is shown as in Figure~1,
and the squares represent the two clusters from 
Rosati \etal~\protect\markcite{rosati99}(1999).
The three panels are for the models shown in
Figure~2:
a) no-evolution model for a 10~Gyr old stellar population,
b) formation at a redshift of 2 followed by passive evolution,
and 
c) as for b) but with a formation redshift of 5.
\par
\end{minipage}
\end{center}
\vspace{3mm}


\vspace{2mm}
\begin{center}
%
\plotfiddle{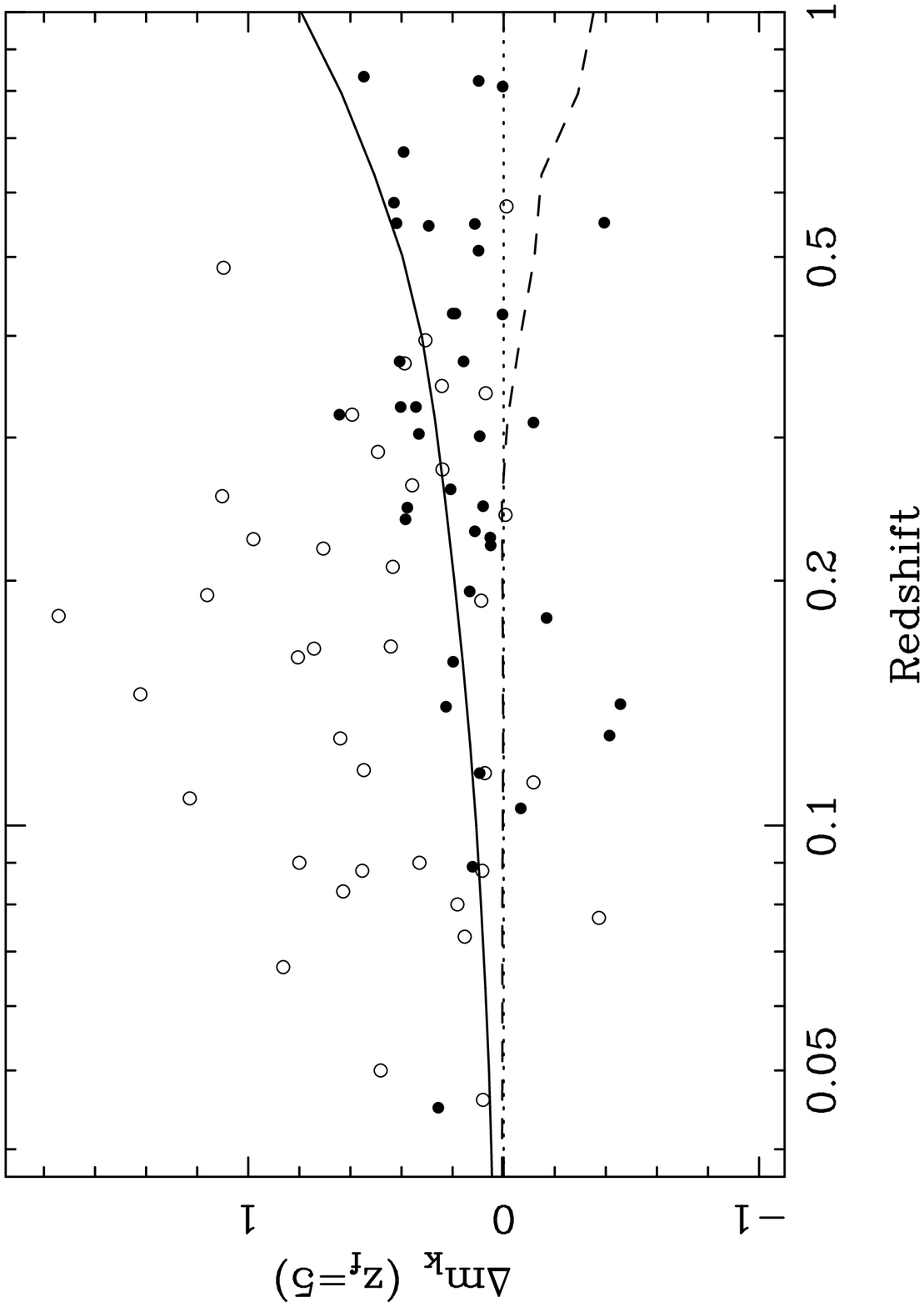}{2.1in}{-90}{32}{32}{-120}{190}
\begin{minipage}{8.75cm}
\small\parindent=3.5mm
{\sc Fig.}~4.---Residuals about the $\zf=5$ model for the X-ray selected BCG sample.
Filled and open symbols indicate BCGs in high- and low-\lx\
clusters respectively.
The lines show the expected locus of the residuals for the three
models shown in Figure~2.
\par
\end{minipage}
\end{center}
\vspace{3mm}

To quantify the amount of evolution allowed by the data, the same
parametric form as employed by
\abkmain---namely $M(z) = M(0) \times (1+z)^\gamma$---was used to estimate 
the growth in the stellar mass content of BCGs.  
Fitting for both $\gamma$ and $M(0)$ indicates that, in the
high-\lx\ sample 
(which best approximates the cluster selection adopted for
the \semianalytic\ models), 
the typical BCG mass has increased by a factor
of $1.9\pm0.3$ (for $\zf=2$) or $1.3\pm0.2$ ($\zf=5$) 
between $z=1$ and the present.
These growth factors are substantially lower than either 
the factor of $\sim$4--5 predicted by the \semianalytic\ models,
or the measured values of 4.6 ($\zf=2$) and 3.2 ($\zf=5$),
of \abkmain.
The growth factor can also be estimated by fitting for $\gamma$ alone if one
assumes a low-redshift normalisation for the model predictions.
However, this currently involves applying a \colour-correction
to low-redshift optical BCG data,
which introduces further uncertainty: 
applying a single $R-K$ correction to the \xray-selected
sample of \lynam\ changes the measured growth factor 
of the high-\lx\ sample by less than 20\%,
whilst using the normalisation adopted by \abkmain---based
on an optically-selected sample---increases the
growth factor by 50\%.
K-band observations of the \lynam\ 
sample are being obtained to circumvent this problem in future
work.

\section{CONCLUSION}
\label{conclusion}

The K-band luminosities of BCGs are correlated with their environment:
clusters with a high \xray\ luminosity contain
BCGs which are brighter, and have a smaller scatter, 
than those BCGs in clusters with a low \xray\ luminosity.
The BCG evolution seen by \abkmain\ has been shown to be
an artifact of a selection bias in their cluster sample;
at high redshifts, their clusters are systematically less \xray\ luminous
than their low-redshift 
sample, and so their BCGs are systematically fainter.

Under the assumption of an Einstein-de Sitter universe,
non-parametric tests show that the only significant evidence for
BCG mass evolution over the range $0.05 \leq z \leq 0.83$
occurs when the dominant stellar population formed
relatively recently ($\zf \leq 2.6$).
Using the same parametric form as \abkmain, the masses of BCGs 
in high-\lx\ clusters are found to have, at most, doubled since $z=1$,
compared to the factor of $\sim 4$ increase predicted, for BCGs in
massive clusters, by the \semianalytic\ models discussed
by \abkmain.

\acknowledgements

DJB acknowledges support from PPARC grant 
GR/L21402
and SAO contract SV4-64008
and RGM that from 
PPARC at Imperial College and Edinburgh.
DJB would like to thank
Peter Draper, Tim Hawarden, and Sandy Leggett for useful discussions.
We thank the referee, Alfonso Arag\'{o}n-Salamanca, for
useful comments that improved the paper,
the service \programmes\ of both UKIRT and IRTF for obtaining
some of the data presented here,
and Piero Rosati and collaborators for providing aperture
magnitudes for the two Lynx clusters.
The United Kingdom Infrared Telescope is operated by the 
Joint Astronomy Centre on behalf of the U.K. Particle Physics and 
Astronomy Research Council.



\begin{references}
\reference{abk98}
Arag\'{o}n-Salamanca, A., Baugh, C. M., \& Kauffmann, G.  1998, \mnras, 297, 427 (ABK98)
\reference{aecc93}
Arag\'{o}n-Salamanca, A., Ellis, R. S., Couch, W. J., Carter, D. 1993, \mnras, 262, 764
\reference{bcf96}
Baugh, C. M., Cole, S., \& Frenk, C. S. 1996, \mnras, 282, L27
\reference{bershady95}
Bershady, M. A. 1995, \aj, 109, 87
\reference{bc}
Bruzual A., G. \& Charlot, S. 1993, \apj, 405, 538   
\reference{b97}
Burke, D. J., Collins, C. A., Sharples, R. M., Romer, A. K., Holden, B. P.,
\& Nichol, R. C. 1997, \apjl, 488, L83
\reference{cm98}
Collins, C. A., \& Mann, R. G. 1998, \mnras, 297, 128 (CM98)
\reference{ukirt-fs}
Casali, M. M., \& Hawarden, T. G. 1992, 
The JCMT-UKIRT Newsletter, No. 3, 33
\reference{castander94}
Castander, F. J., Ellis, R. S., Frenk, C. S., Dressler, A.,  
\& Gunn, J. E. 1994, \apjl, 424, L79 
\reference{depropris99}
De Propris, R., Stanford, S. A., Eisenhardt, P. R., Dickinson, M.,
\& Elston, R. 1999, \aj, 118, 719
\reference{dressler78}
Dressler, A. 1978, \apj, 222, 23
\reference{els62}
Eggen, O. J., Lynden-Bell, D., \& Sandage, A. R. 1962, \apj, 136, 748
\reference{ellis97}
Ellis, R.S., Smail, I., Dressler, A., Couch, W.J., Oemler, A., Butcher, H., \&
Sharples, R.M., 1997, \apj, 483, 582
\reference{gl-94}
Gioia, I. M., \& Luppino, G. A. 1994, \apjs, 94, 583
\reference{hall98}
Hall, P. B., Green, R. F., \& Cohen, M.  1998, \apjs, 119, 1 
\reference{h97}
Henry, J. P., et al. 1997, \aj, 114, 1293 
\reference{h99}
Henry, J. P. 1999, \apj, submitted
\reference{holden97}
Holden, B. P., Romer, A. K., Nichol, R. C., \& Ulmer, M. P. 1997, \aj, 114, 1701 
\reference{jimenez99}
Jimenez, R., Friaca, A., Dunlop, J., Terlevich, R., Peacock, J.,
\& Nolan, L. 1999, \mnras, 305, L16
\reference{kc98}
Kauffmann, G., \& Charlot, S. 1998, \mnras, 294, 705
\reference{kw93}
Kauffmann, G., \& White, S. D. M. 1993, \mnras, 264, 201
\reference{larson69}
Larson, R. B. 1969, \mnras, 145, 405
\reference{pdl99}
Lynam, P. D., Collins, C. A., James, P. A., B\"{o}hringer, H., \&
Neumann, D. M. 1999, preprint (astro-ph/9908348)
\reference{madau98}
Madau, P., Pozzetti, L., \& Dickinson, M. E. 1998, \apj, 498, 106
\reference{n97}
Nichol, R. C., Holden, B. P., Romer, A. K., Ulmer, M. P.,
Burke, D. J., \& Collins, C. A. 1997, \apj, 481, 644
\reference{opl98}
Oke, J. B., Postman, M.,  \& Lubin, L. M. 1998, \aj, 116, 549 
\reference{morphs99}
Poggianti, B. M., Smail, I.,  Dressler, A.,  Couch, W. J., Barger, A. J., 
Butcher, H., Ellis, R. S., \& Oemler, A., Jr. 1999, \apj, 518, 576 
\reference{bsharc}
Romer, A. K., et al. 2000, \apjs, in print
\reference{rosati99}
Rosati, P., Stanford, S. A., Eisenhardt, P. R., Elston, R.,  Spinrad, H., 
Stern, D., \& Dey, A. 1999, \aj, 118, 76 
\reference{schlegel98}
Schlegel, D., Finkbeiner, D., \& Davis, M. 1998, \apj, 500, 525
\reference{sed95}
Stanford, S. A., Eisenhardt, P. R., \& Dickinson, M. 1995, \apj, 450, 512 
\reference{sed98}				     
Stanford, S. A., Eisenhardt, P. R., \& Dickinson, M. 1998, \apj, 492, 461
\reference{tb94}
Thimm, G. J., \& Belloni, P. 1994, \aap, 289, L27 
\reference{v96}
van Dokkum, P. G., \& Franx, M., 1996, \mnras, 281, 985
\reference{vf98}
van Dokkum, P. G., Franx, M., Kelson, D. D., \& Illingworth, G. D. 1998,
\apjl, 504, L17
\reference{v99}
van Dokkum, P. G., Franx, M., Fabricant, D., Kelson, D. D., \& Illingworth, G. D. 1999,
\apjl, 520, L95
\end{references}
\end{document}